\documentclass[journal]{IEEEtran}
\usepackage{graphicx}
\usepackage{booktabs}
\usepackage{tabularx}
\usepackage{amsmath}
\usepackage{booktabs}
\usepackage{caption}
\usepackage{multirow}
\usepackage{adjustbox}
\usepackage{orcidlink}
\usepackage{subfigure}
\usepackage[nospace,noadjust]{cite}

\ifCLASSINFOpdf
\else
\fi
\begin{document}
\bstctlcite{no_dashes}
\title{Development  
of Immersive Virtual and
Augmented Reality-Based Joint
Attention Training Platform for
Children with Autism}
\author{Ashirbad Samantaray\orcidlink{0009-0003-6121-4205}\thanks{
This work was supported by the Ministry of Health and Family Welfare, Government of India. (Ashirbad Samantaray and Taranjit Kaur contributed equally to this work.) (Corresponding authors: Sheffali Gulati; Tapan Kumar Gandhi.)}
\thanks{This work involved human subjects in its research. Approval of all ethical and experimental procedures and protocols was granted by the Institute Ethics Committee, All India Institute of Medical Sciences, New Delhi, under Application No. EC-178/06.04.2018 and OP-12/13.01.2023.}
\thanks{
Ashirbad Samantaray, Sapna S Mishra, Kritika Lohia, Chayan Majumder and Tapan Kumar Gandhi are with the Department of Electrical Engineering, Indian Institute of Technology Delhi, New Delhi, India (email: samantaray.ashirbad@gmail.com; eez208443@iitd.ac.in; kritika.lohia@ee.iitd.ac.in; chinsi@gmail.com; tapan.kumar.gandhi@ee.iitd.ac.in).}, Taranjit Kaur\orcidlink{0000-0001-5972-3957}\thanks{Taranjit Kaur is with the School of Artificial Intelligence and Data Sciences, Indian Institute of Technology Jodhpur, Rajasthan, India (e-mail: taranjit@iitj.ac.in).}, Sapna S Mishra \orcidlink{0000-0002-5304-7131}, Kritika Lohia \orcidlink{0000-0002-6568-2714}, Chayan Majumder\orcidlink{0009-0003-6328-328X}, Sheffali Gulati\orcidlink{0000-0003-1439-9959}\thanks{Sheffali Gulati is with the Department of Pediatrics, All India Institute of Medical Sciences Delhi,
New Delhi, India (e-mail: sheffaligulati@gmail.com).}, Tapan Kumar Gandhi \orcidlink{0000-0002-3532-9389}
}


\maketitle
\begin{abstract}

Joint Attention (JA), a crucial social skill for developing shared focus, is often impaired in children with Autism Spectrum Disorder (ASD), affecting social communication and highlighting the need for early intervention. Addressing gaps in prior research, such as limited use of immersive technology and reliance on distracting peripherals, we developed a novel JA training platform using Augmented Reality (AR) and Virtual Reality (VR) devices. The platform integrates eye gaze-based interactions to ensure participants undivided attention.  To validate the platform, we conducted experiments on ASD (N=19) and Neurotypical (NT) (N=13) participants under a trained pediatric neurologist’s supervision. For quantitative analysis, we employed key measures such as the number of correct responses, the duration of establishing eye contact (s), and the duration of registering a response (s), along with correlations to CARS scores and age. Results from AR-based experiments showed NT participants registered responses significantly faster (p\textless0.00001) than ASD participants. A correlation (Spearman coefficient=0.57, p=0.03) was found between ASD participants’ response time and CARS scores. A similar trend was observed in VR-based experiments. When comparing response accuracy in ASD participants across platforms, AR yielded a higher correctness rate (92.30\%) than VR (69.49\%), indicating AR’s greater effectiveness.  These findings suggest that immersive technology can aid JA training in ASD. Future studies should explore long-term benefits and real-world applicability.
\end{abstract}


\section{Introduction}
Autism Spectrum Disorder (ASD) is a neurodevelopmental condition characterized by atypical developmental trajectories and the presence of restricted, repetitive behaviors \cite{lord2018autism}. It is typically identified in early childhood, most often between one and three years of age \cite{ousley2014autism}. Recent global estimates suggest that ASD affects approximately 1 in 36 children \cite{maenner2023prevalence}. In the Indian context, a population-based study by Arora et al. reported prevalence rates as high as 1 in 125 among children aged 2–6 years and 1 in 80 among those aged 6–9 years, with an overall prevalence of 1 in 89 across surveyed regions \cite{arora2018neurodevelopmental}. ASD significantly impacts both emotional and cognitive functioning, thereby limiting social communication and developmental progress \cite{de1998parental}. Early manifestations of the disorder commonly include abnormalities in speech and language acquisition \cite{tager2005language}, engagement in repetitive behaviors \cite{turner1999annotation}, diminished responsiveness to social cues \cite{schneider2015reduced}, atypical gaze patterns \cite{nakano2010atypical}, and marked difficulties in establishing joint attention \cite{mundy1989theoretical,mundy1994joint}.

Joint Attention (JA) refers to the ability of two or more individuals to focus their attention on an object or person \cite{tomasello1986joint}. This shared attentional process plays a pivotal role in the early stages of language acquisition and social learning \cite{whalen2006collateral}, as it enables individuals to interpret their environment and infer the mental states of others. Previous research has consistently established JA as an essential precursor to developing linguistic skills \cite{mundy1990longitudinal,bruner1975ontogenesis,rollins1998intervention}. Importantly, impairments in JA during infancy have been strongly associated with an increased likelihood of developing autism \cite{mundy1989theoretical,mundy1994joint}. Moreover, converging evidence indicates that an infant's JA ability is closely linked to their broader information processing capacity \cite{striano2006joint,mundy2010infant,kaplan2006challenges,mundy2018review}. Mundy et al. \cite{mundy1994joint} further demonstrated that the extent to which an infant can successfully influence another individual’s attention toward an event or object may serve as an indicator of autism severity.

Children diagnosed with ASD typically experience delays in the development and acquisition of JA skills when compared to their Neurotypical (NT) counterparts \cite{carpenter2002interrelations}. This delay can include difficulty initiating JA through actions such as, finger-pointing or shifting eye gaze, and challenges in responding to JA cues provided by parents, guardians, or caregivers \cite{stone1997nonverbal}. Nevertheless, prior research suggests that early interventions can help individuals with ASD to improve their communication skills \cite{kasari2012longitudinal,whalen2006collateral}. In the absence of such rehabilitative measures, individuals can face persistent challenges in establishing social relationships and engaging in effective communication. However, it is crucial to acknowledge that the efficacy of JA intervention techniques can vary between individuals. This variability is influenced by a range of factors, including social conditioning and stages of individual development \cite{bono2004relations}. 

Given these challenges, there is a growing need for early interventions and targeted JA skill training for children with ASD. Over the years, a variety of intervention methods have been developed to support the early acquisition of JA skills, ranging from traditional clinician or caregiver-mediated approaches to technology-assisted strategies \cite{meindl2011initiating,rocha2007effectiveness,kumazaki2018impact,yazdanian2025virtual}. With the rapid advancement of computational power, immersive modalities such as Virtual Reality (VR) and Augmented Reality (AR) are increasingly being investigated as innovative platforms for JA training. In particular, most of these interventions are based on behavioral principles \cite{ravindran2019virtual}, particularly those derived from Applied Behavior Analysis (ABA), which provide the theoretical and empirical foundation for systematically shaping and reinforcing JA behaviors. 

ABA-based methods involve intensifying and repeating cues to achieve the desired response while decreasing unwanted behaviors \cite{kirkham2017line,grey2005evaluating}. Discrete Trial Training (DTT) \cite{delprato2001comparisons}, a widely adopted ABA-derived technique, employs positive reinforcement to strengthen desired behavior while correcting for errors through feedback and task-mediated prompts \cite{smith2015evidence}. Such behavioral training paradigms constitute the foundation of various existing JA training programs \cite{murza2016joint}\cite{roane2016applied}.

Traditionally, JA training has been conducted through in-person interventions, in which a trained practitioner works directly with the child using physical objects to elicit and reinforce JA behaviors \cite{white2011best,whalen2003joint,tomasello1986joint}. Although effective, this approach is resource-intensive, requiring substantial time and effort from practitioners, and presents challenges in ensuring consistency, particularly in the timing and delivery of cues. With advances in digital technologies, there has been a marked increase in the development and deployment of technology-mediated JA interventions. These approaches frequently employ virtual environments with avatars that deliver standardized JA cues \cite{amat2021design,jyoti2019virtual}. Virtual platforms offer key advantages: they allow consistent replication of training exercises, ensure precise delivery of cues, and reduce reliance on human resources, thereby improving efficiency relative to traditional interventions. Despite these advances, significant gaps remain in the application of immersive technologies for JA training in children with ASD \cite{yazdanian2023virtual}. Prior studies \cite{jyoti2019virtual,amat2021design} have largely relied on screen-based paradigms, with limited integration of eye-tracking capabilities. Moreover, systematic comparisons across immersive platforms such as Virtual Reality (VR) and Augmented Reality (AR) remain scarce, highlighting the need for further investigation.

In this paper, we introduce an innovative immersive JA training platform that employs virtual avatars as task mediators and is compatible with both VR and AR headsets equipped with integrated eye trackers. The platform is uniquely designed to rely exclusively on participants’ eye gaze for interaction, thereby enabling precise, real-time capture of gaze-based behaviors through head-mounted displays (HMDs). This capability allows the application to accurately record and evaluate task performance during training sessions. The present work represents the culmination of our efforts in developing and validating this platform across immersive VR and AR environments. The overarching aim is to demonstrate the potential of this technology-driven approach to enhance JA skill training. Specifically, we evaluate the feasibility and acceptance of the system among individuals with ASD, as well as its efficacy in supporting the acquisition of JA skills.

\section{Related Works}
In recent years, numerous studies \cite{imai2003physical,kumazaki2018impact,white2011best,whalen2003joint,little2016gaze} have explored methods to help children with ASD develop JA, using robots, virtual avatars, and human-mediated interaction. Immersive technologies such as VR and AR have advanced autism care \cite{bozgeyikli2017survey}, offering realistic, cost-effective, and engaging 3D environments via head-mounted displays (HMDs) \cite{cheng2015using}. Their heightened immersion can enhance learning and support skill training; prior work indicates strong acceptance of HMD-based rehabilitation among individuals with ASD \cite{lahiri2012design,buono2016proceedings}. These platforms deliver controlled digital environments for diverse rehabilitation programs \cite{newbutt2016brief,bradley2018autism}.

For JA training, Billing et al. \cite{billing2020dream} analyzed behavioral and physiological responses during Robot-Assisted Therapy (RAT) \cite{thill2012robot,belpaeme2018social}, reporting positive outcomes. Caruana et al. \cite{caruana2018joint}, Jyoti et al. \cite{jyoti2019virtual,jyoti2019conf,jyoti2022portable,jyoti2020design,jyoti2020human}, and Little et al. \cite{little2016gaze} used screen-based eye tracking, and other work has employed virtual avatars in VR. Ravindran et al. \cite{ravindran2019virtual} built a low-cost phone-based VR solution; Amaral et al. \cite{amaral2017novel} integrated EEG with immersive VR; Mei et al. \cite{mei2018towards} studied high-functioning ASD participants; and Perez-Fuster et al. \cite{perez2022enhancing} used AR, showing that targeted AR interventions can improve JA.

A systematic review by Yazdanian et al. \cite{yazdaninreview} highlights limited use of immersive VR/AR for JA training and few systems with real-time eye tracking, and examines the potential and acceptance of such approaches. To the best of our knowledge, this is the first AR-based, avatar-mediated JA training platform that integrates eye tracking for participant interaction.

\section{Materials and Methodology}
In this study, we employ two distinct immersive hardware configurations to evaluate the proposed JA training application. Each configuration consists of a head-mounted display (HMD) with an integrated eye tracker, enabling participants to interact exclusively through gaze-based input. The first setup utilizes a FOVE 0 VR headset, while the second employs a Microsoft HoloLens 2 AR headset. Detailed specifications of the hardware and software components used in both setups are provided in the following sections.

\subsection{VR Setup (FOVE 0 Headset)}
The experimental apparatus for the VR-based JA training platform includes a FOVE 0 VR headset and a Dell Alienware m15-r3 gaming laptop operated by the experimenter. The FOVE 0 headset, developed by Fove Inc., features a 2560 × 1440-pixel resolution display, a 100° field of view, and a six-degree-of-freedom tracking system. It renders at a stable frame rate of 70 frames per second and incorporates a built-in stereo infrared eye tracker, enabling participants to interact with the JA application in real time. The experimenter’s laptop features a 15.6-inch Full HD display and is configured to run the VR application. It is powered by an Intel Core i7-10750H processor with 16 GB DDR4 266 MHz RAM and an Nvidia GeForce RTX 2070 graphics card with 8 GB GDDR6 RAM, providing the computational resources necessary for smooth execution of the JA application. Operating on Windows 10 Pro, the laptop runs the training application, which is rendered on the VR headset via a wired connection.

\begin{figure}[h]
    \centering
\includegraphics[width=8.8cm,height=6cm]{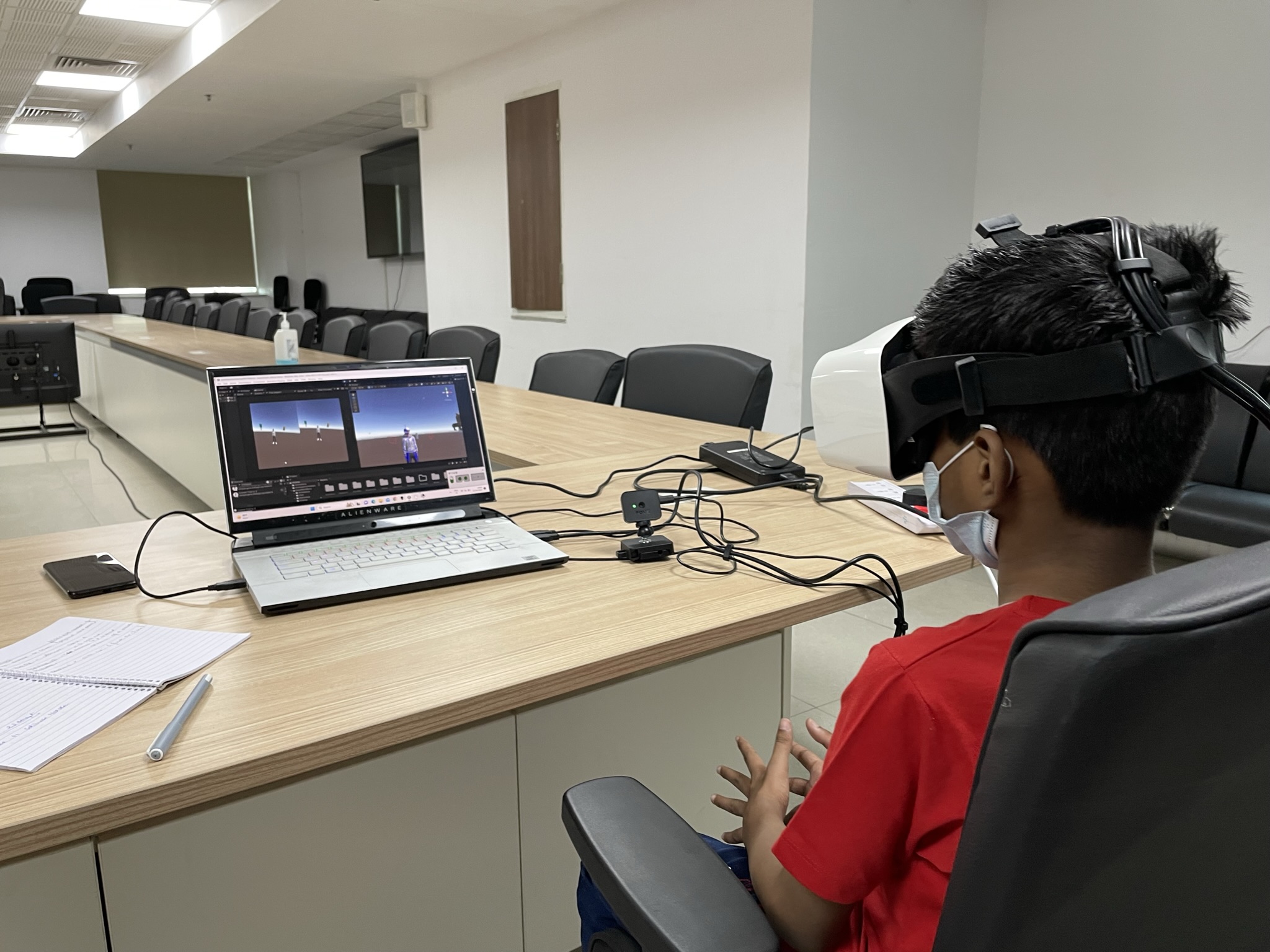}
    \caption{Participant wearing FOVE 0 VR headset for JA training. }
    \label{fig:vr_participant_setup}
\end{figure}

\subsection{AR Setup (HoloLens 2 Headset)}
The AR setup includes a HoloLens 2 headset and a laptop (used by the experimenter). The JA training application is installed on the AR headset and runs directly on it. 
The headset, developed by Microsoft Inc., features a 2K resolution display for a high-quality visual experience. It also includes a 1-megapixel time-of-flight sensor to enhance spatial awareness and precision. The headset has an 8-megapixel camera, accelerometer, gyroscope, and magnetometer for 6 degrees of freedom monitoring and spatial mapping. The HoloLens 2 is powered by a Qualcomm Snapdragon 850 processor and a custom-built Holographic Processing Unit. It has 4GB LPDDR4x system DRAM and 64GB UFS 2.1 storage. Additionally, it also features an internal microphone and speaker. Flexible communication and data transfer are made possible through Wi-Fi 5, Bluetooth 5, and USB Type-C connectivity options. The headset utilizes four visible light camera sensors for head tracking and two infrared camera sensors for accurate eye tracking. In this setup, the experimenter utilized the same laptop that was employed for the VR configuration.

\begin{figure}[h!]
    \centering
    \includegraphics[width=8.8cm,height=6cm]{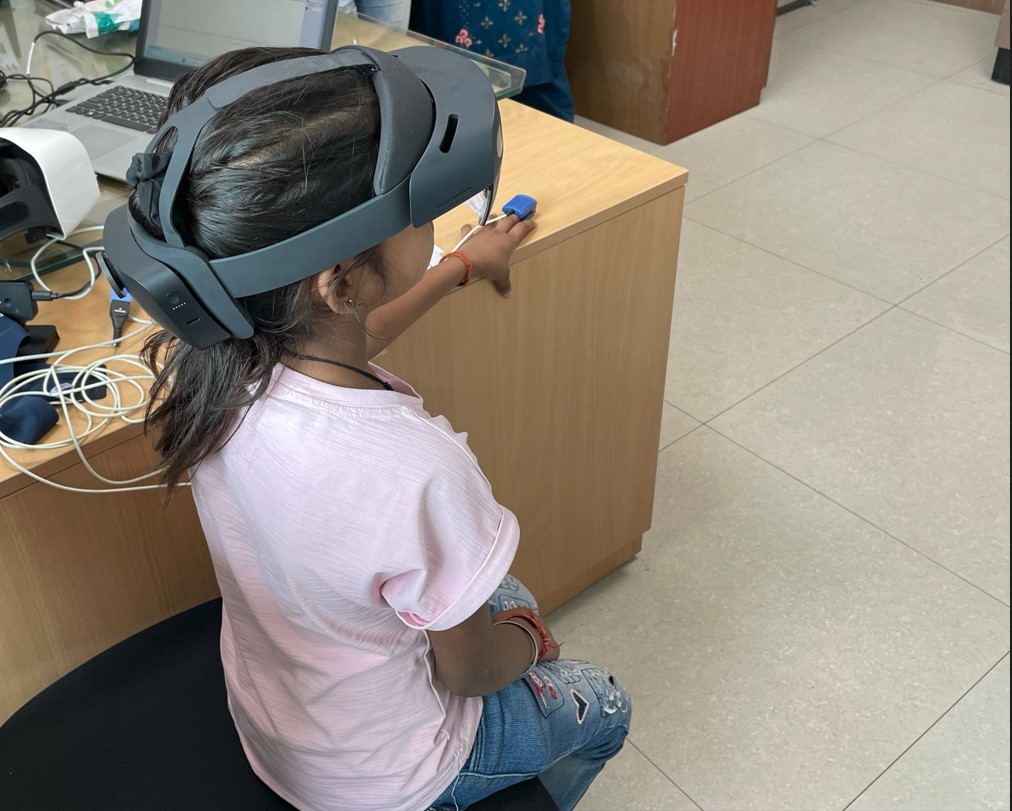}
    \caption{Participant wearing HoloLens 2 AR headset for JA training}
    \label{fig:ar_participant_setup}
\end{figure}

\subsection{Software}

\begin{figure} [h!]
    \centering
    \includegraphics [width=9cm, height=6cm]{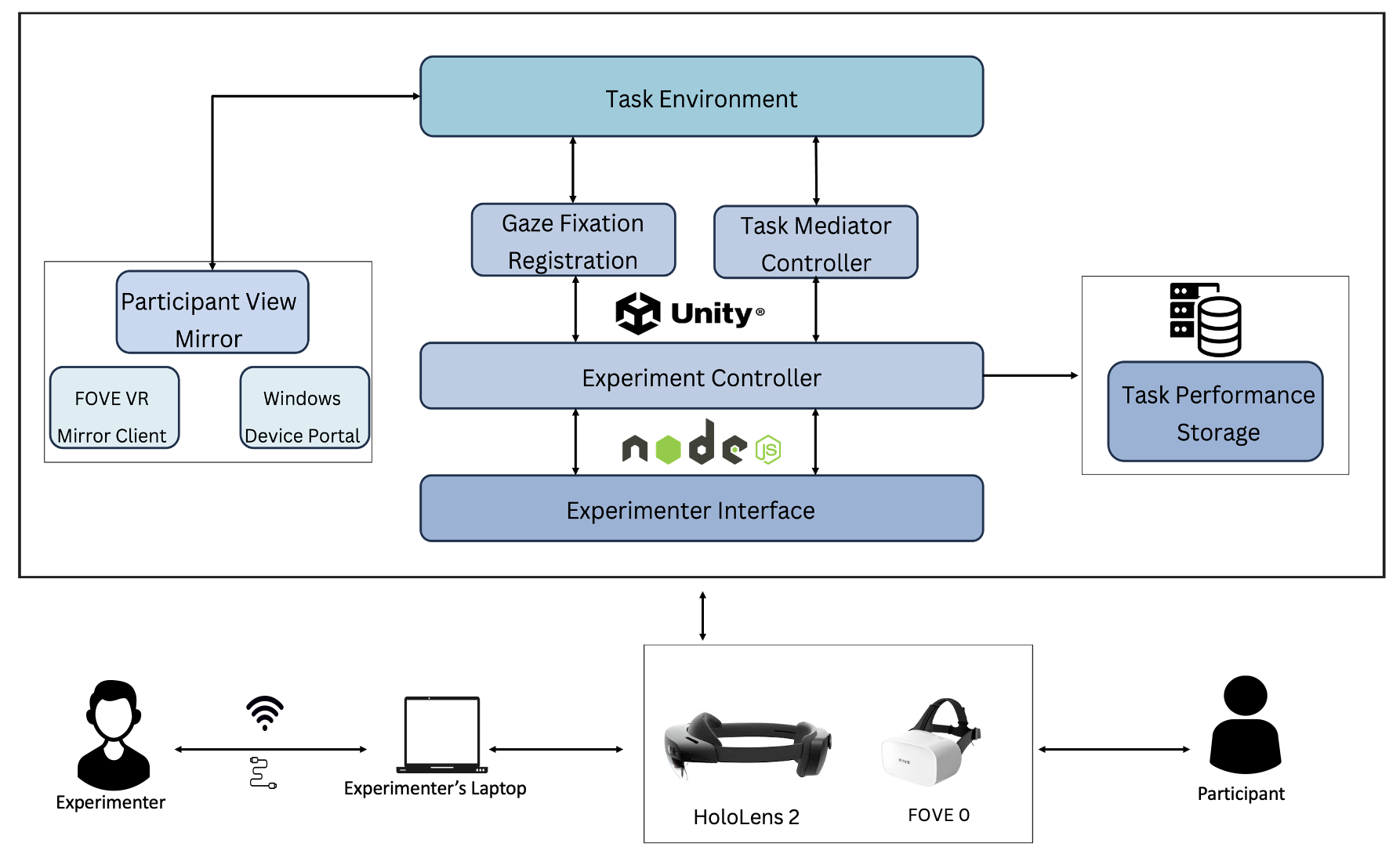}
    \caption{System Design Overview of JA Training Platform. The task environment designed in Unity delivers JA exercises with modules for gaze fixation registration and task mediation. An experiment controller developed using Node.js coordinates data flow between the experimenter interface. It logs task performance and mirrors the participant's view for monitoring. Hardware components include an experimenter's laptop, a Microsoft HoloLens 2 AR device, and a FOVE 0 VR device.}
    \label{fig:combined_system_design}
\end{figure}

The JA training application is developed using the Unity 3D game engine (v2021.3.12f1) \cite{Unity2021}, with C\# as the primary programming language. Interaction with VR and AR devices is enabled through the FOVE \cite{FOVE} and Mixed Reality Toolkit (MRTK) \cite{MRTK} SDKs. The stimulus environment features a virtual avatar acting as a task mediator, delivering cues toward 3D objects within the scene. The avatar is created using the Ready Player Me platform \cite{ReadyPlayerMe} and refined in Blender \cite{Blender}. Animations for finger-pointing and head-turning cues are generated using Mixamo \cite{Mixamo} and further adjusted in Unity with the Animation Rigging package \cite{UnityAnimationRigging}. The final task environment incorporates seven 3D objects obtained from the Unity Asset Store \cite{UnityAssetStore}.

\begin{figure*} [htpb]
    \centering
    \includegraphics[width=1\textwidth]{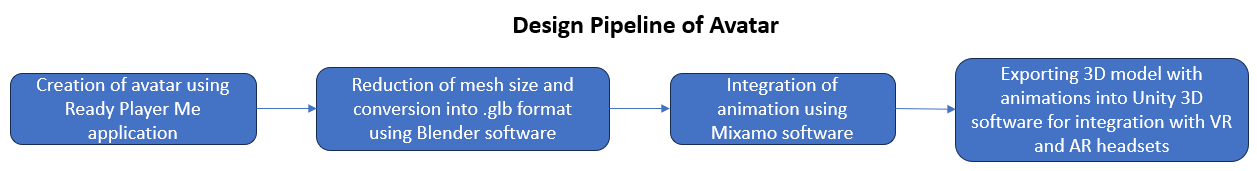}
    
    \caption{Process pipeline for the design and development of an avatar as a task mediator in our JA application}
    \label{fig:avatar_design_pipeline}
\end{figure*}

The JA training application comprises of separate modules for both the VR and AR setups. However, it is to be noted that most of the modules are common throughout both setups. The \textit{JA Task Environment} is responsible for rendering the developed stimulus into the VR and AR headset, including the placement and position of the virtual avatar and 3D models. This module is configured to be compatible with both headsets. 

The \textit{Task Mediator Controller} manages the cue delivery animations of the avatar. We have utilized a Posner style cueing paradigm \cite{posner1980orienting} to observe the shift of visual attention of autistic children after the delivery of various types of cues. This module selects two 3D objects at random and places them on either side of the avatar. Further, it randomly selects a single object as the target and executes animations on the avatar based on the experiment settings. 

The \textit{Gaze Fixation Registration} is responsible for tracking the participant’s gaze and managing its interaction with our JA training application in real-time. This module achieves this by accessing the eye trackers of FOVE 0 and HoloLens 2 headsets. The \textit{Experimenter Interface} is responsible for providing the experimental parameters as input to the JA training application. The experimenter enters the parameters at the start of each session and provides feedback at the end of the session through a user interface.

The \textit{Participant View Mirror} enables the experimenter to observe what the participant watches in real-time while interacting with our JA training application. For the VR setup, this is facilitated by the \textit{FOVE VR Mirror Client} software through a wired connection between the experimenter's computer and the VR headset. In the case of AR setups, this is achieved using the \textit{Windows Device Portal} through a wireless connection between the experimenter’s computer and the AR headset. 

The \textit{Task Performance Storage} module is responsible for recording the task performance metrics of the participants. This module operates on the experimenter's computer within the AR setup, while the remaining parts of the application run independently on the HoloLens 2 AR headset. A middleware server is configured on the experimenter's computer to establish communication with the AR headset, and it is developed using the Node.js framework. To retrieve the participant performance data, it sends HTTP web requests to an API endpoint of the JA application running on the AR headset. Upon receiving the data, it is parsed and stored in JSON format for subsequent analysis.

Finally, the \textit{Experiment Controller} coordinates the interactions between various modules. This module actively monitors the user's activity and passes the information to the concerned modules accordingly. It is responsible for assessing the correctness of the participant's response, which is a fundamental metric for evaluating the effectiveness of the JA training exercises. It also verifies the timing and duration of the participant's responses and sends them to the \textit{Task Performance Storage} module to log them. This module also monitors the prolonged periods of inactivity from the participant's side, which, upon identification, terminates the experiment session. 

\section{Experiment Design}
\subsection{Experiment Protocol}

We conduct a feasibility study to evaluate the effectiveness of the VR- and AR-based JA training platforms for participants with ASD. To assess the acceptability of the applications, we compare task performance metrics between ASD participants and age-matched neurotypical (NT) controls, as well as across the VR and AR configurations.
All participants have no prior experience with VR or AR devices. Each participant is first evaluated by a trained pediatric psychologist, who categorizes them into ASD or NT groups using standardized assessment scales. Following the evaluation, participants are seated comfortably in a well-lit room and are informed of their right to withdraw at any point if they experience discomfort. Once the participant confirms readiness, the experimenter mounts the VR or AR headset.

The experimenter then configures the application with fixed experimental parameters across both setups: (a) minimum gaze duration on the avatar to register eye contact (2 seconds), (b) minimum fixation duration on the target object to register a response (2 seconds), (c) duration of avatar cue delivery animation (5 seconds), and (d) number of trials per session (2). Prior to beginning the tasks, participants are shown a video demonstration and provided with a verbal description of their expected task.

To mitigate VR-induced discomfort, the total session duration is kept below 20 minutes. The VR-based experiment is conducted first, followed by the AR-based study, with participants completing the same tasks in both configurations. Additionally, the application includes an automated feature that terminates the session if no participant activity is detected for 20 minutes.
\subsection{Task Description}
\begin{figure} [h]
    \centering
    \includegraphics [width=8cm,height=6cm]{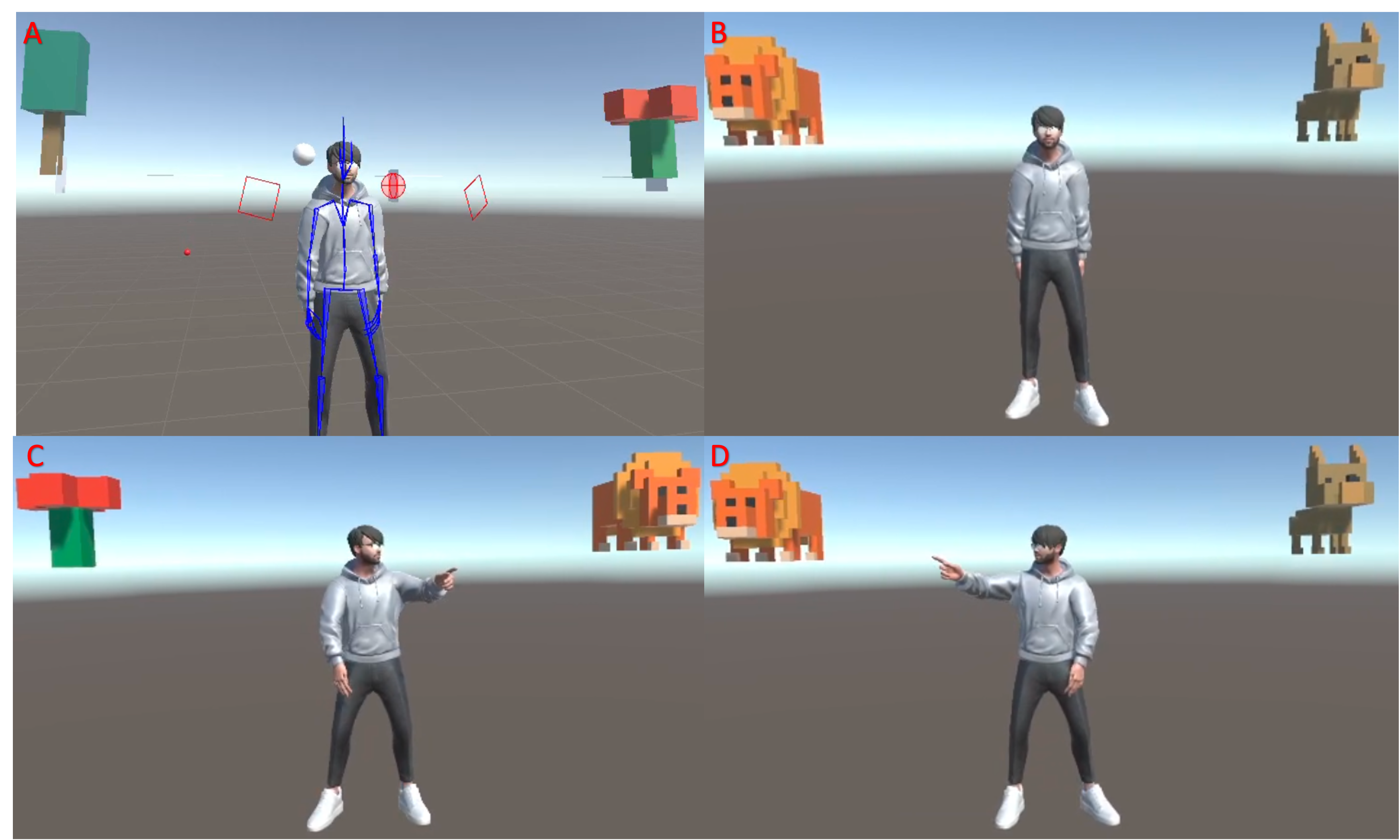}
    \caption{Cue Delivery Sequence of Avatar in VR platform (A) Overview of VR environment in Unity Editor (B) Waiting to establish eye contact (C) Delivery of finger pointing cue by avatar to right and left (D) side}
    \label{fig:vr_cue_sequence_ashirbad_avatar}
\end{figure}

The experiment session starts with calibrating the eye-tracking system to accurately measure the participant's eye gaze fixation. The built-in calibration functions of the FOVE and HoloLens 2 headset are used to complete this step.
Subsequently, the JA training application shows the stimulus to the participant with an avatar placed in the centre and two objects on either side. The participant is first required to gaze at the eye region of the avatar for a minimum duration, as specified in the experiment settings.

\begin{figure} [h]
    \centering
    \includegraphics [width=8cm,height=6cm]{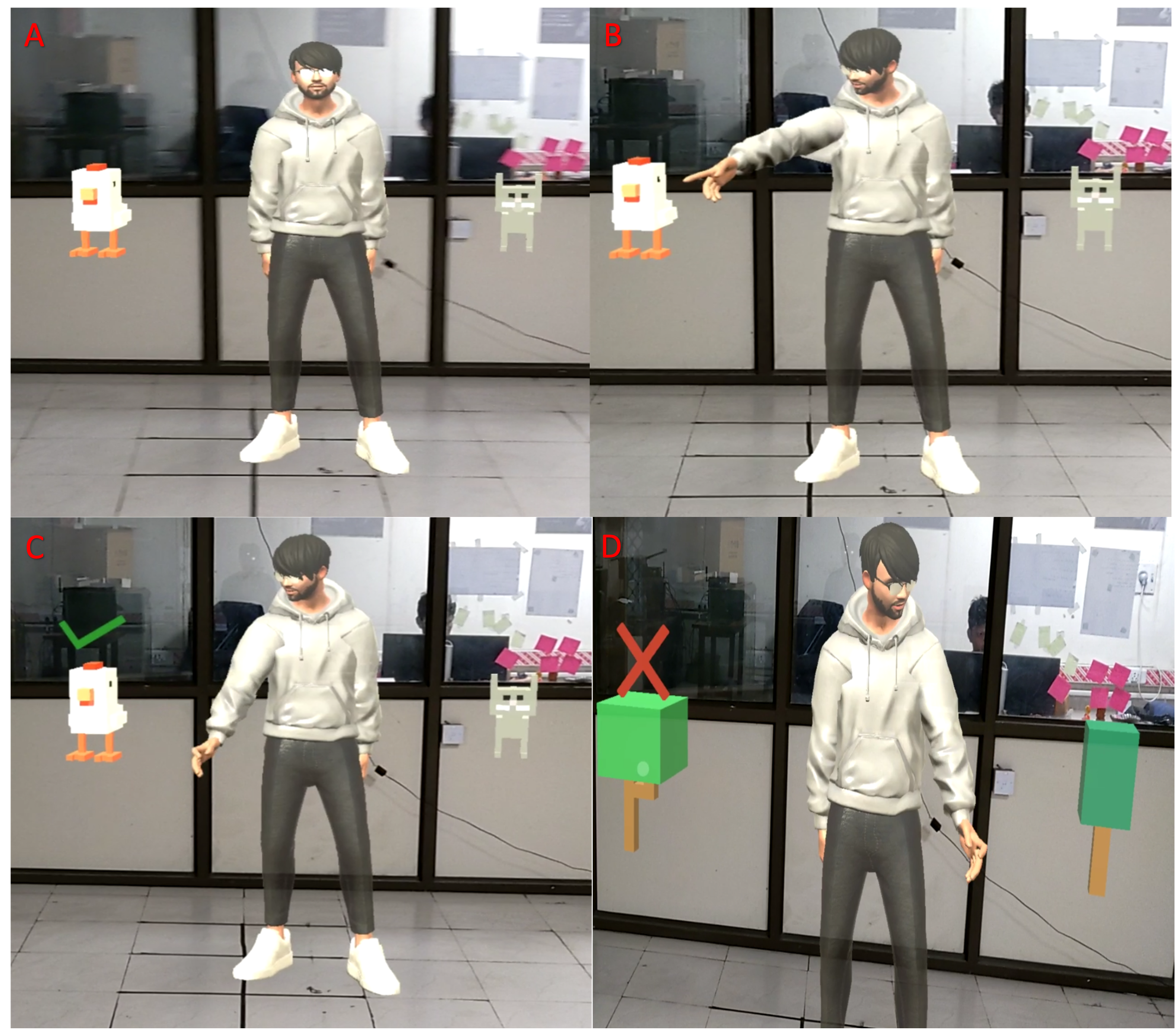}
    \caption{Cue Delivery Sequence of Avatar in AR platform (A) Waiting to establish eye contact (B) Delivery of Cue and waiting for user to register a response on target object (C) Feedback to the participant for correct registration of target object (D) Feedback to the participant for incorrect registration of target object  }
    \label{fig:ar_cue_sequence_ashirbad_avatar}
\end{figure}

Upon establishing eye contact with the avatar, as seen in Fig. \ref{fig:vr_cue_sequence_ashirbad_avatar} and \ref{fig:ar_cue_sequence_ashirbad_avatar}, the application subsequently initiates a sequence of visual cues and animations aimed at directing the participant's attention towards a specified target object. These cues are characterized by an initial head-turning motion, which is then followed by a gesture of pointing with the finger.

Once the cues have been delivered, the participant is required to register a response by gazing at an object for 2 seconds. The application provides visual feedback based on the correctness of the response.

\subsection{Participants}
We recruited a total of 29 participants for our study 
(ASD=16 and NT=13). The age range of the participants was between 6 and 13 years (ASD: mean=9.46, SD=2.27; NT: mean=9.99, SD=2.32).
The volunteers were recruited from the Autism Clinic, Child Neurology Division, Department of Pediatrics, All India Institute of Medical Sciences, Delhi. Participants diagnosed with mild to moderate high-functioning autism and having an Intelligence Quotient (IQ) greater than 80 (high functioning ASD) were included in our study, while participants with any other associated neurological disorders/conditions/comorbidities were excluded. Initially, we had recruited 19 ASD participants who were previously diagnosed with ASD, but upon reevaluation by a pediatric neurologist and child psychologist, it was found that 3 individuals were in remission and therefore excluded from the study. 

\begin{table}[htbp]
  \centering
  \caption{Participants Clinical Characteristics}
  \begin{tabularx}{8cm}{XXX}
    \toprule
             & ASD (n=16) & NT (n=13) \\
    \cmidrule{2-3}
    Participants &  Mean     &  Mean \\
    \midrule
    Age      & $9.46\pm2.27$ & $9.99\pm2.32$ \\
    CARS Score\textsuperscript{1} & $32.03\pm2.87$ & $9.23\pm5.58$ \\
    SCQ Score\textsuperscript{2} & $32.86\pm7.83$ & $11\pm3.65$ \\
    SRS 2\textsuperscript{3} & $66.73\pm2.63$ & $18.46\pm8.72$ \\
    \bottomrule
  \end{tabularx}
  \label{tab:participantDetails}%
  \footnotesize

    \textsuperscript{1}ASD range=28-33.5, \textsuperscript{2}ASD range=15+, \textsuperscript{3}ASD range=60-75
\end{table}   

During the study, all participants were assessed by a licensed psychologist using standardized autism rating scales. The Diagnostic and Statistical Manual of Mental Disorders, Fifth Edition (DSM-5)  \cite{american2013diagnostic} was employed for defining ASD. NT participants were defined as children who did not have any known neurological disorder, with typical development and no behavioral issues. Additionally, the Childhood Autism Rating Scale High Functioning Second Edition (CARS-2) \cite{schopler2010childhood} was used to measure the severity of autism of each participant. Furthermore, the Social Communication Questionnaire Lifetime Score (SCQ-Lifetime) \cite{rutter2003social} and the Social Responsiveness Scale Second Edition (SRS-2) \cite{constantino2012social} were also employed to evaluate the social and behavioral level of each participant. Before conducting the experiments, signed consent was taken from the participant's parents.

\section{Results}
\begin{figure}[h]
    \centering
    \includegraphics[width=8cm,height=5cm]{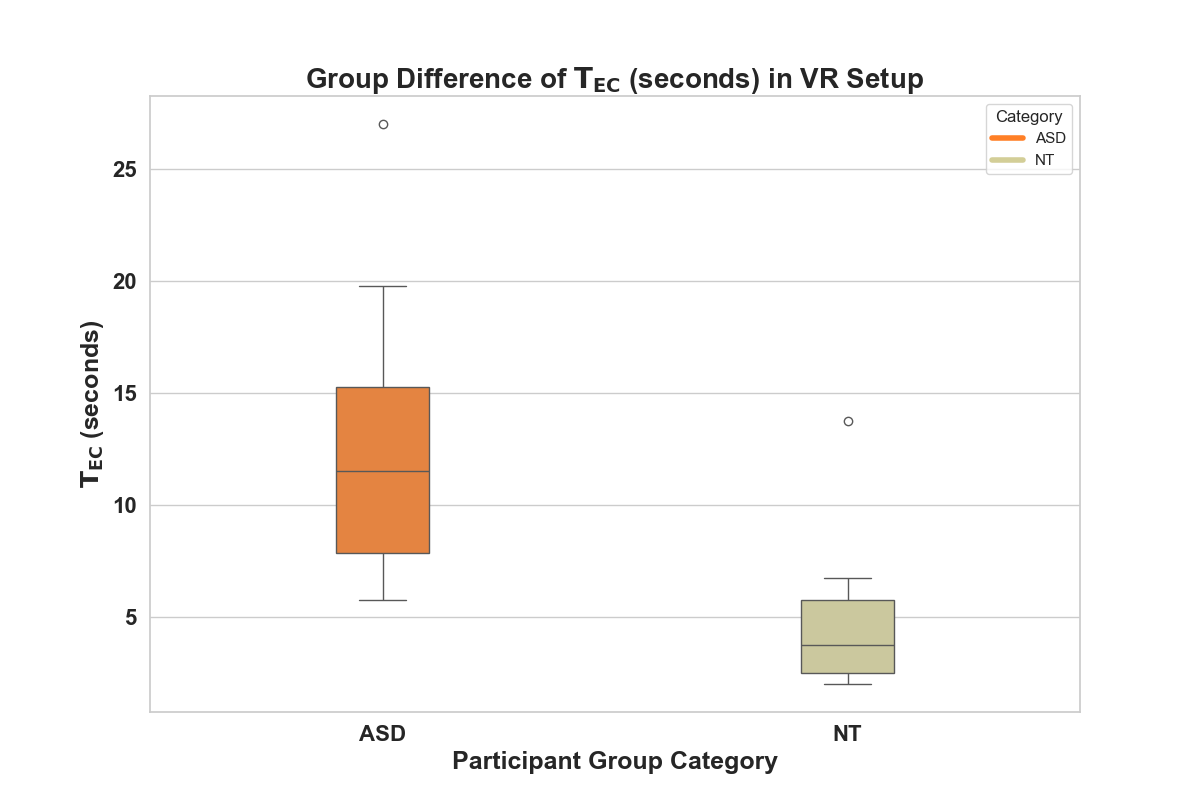}
    \caption{Duration to Establish Eye Contact (T\textsubscript{EC}) comparison between ASD and NT participants (ASD=11.5s, NT=3.75s) (z=4.37,p=0.00007) in VR setup}
    \label{fig:vr_eye_contact_comparison}
\end{figure}
\begin{figure}[h]
    \centering
    \includegraphics[width=8cm,height=5cm]{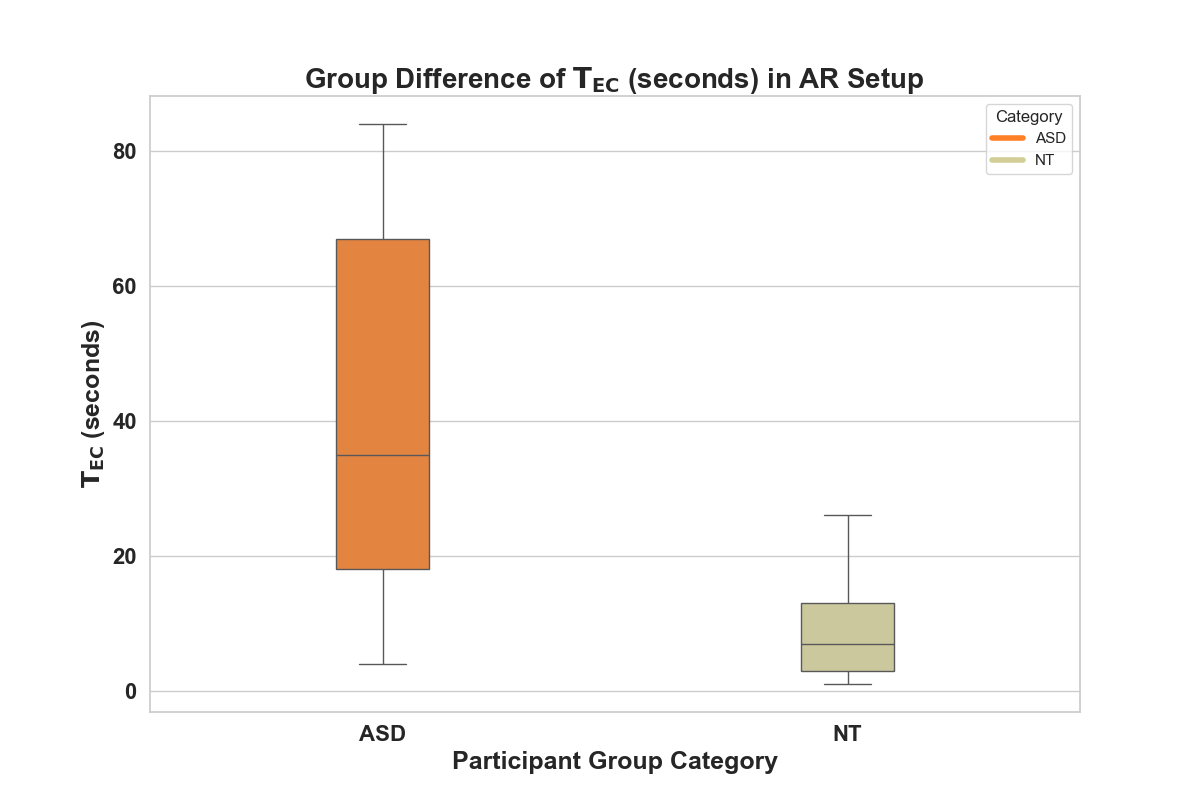}
    \caption{Duration to Establish Eye Contact (T\textsubscript{EC}) comparison between ASD and NT participants (ASD=35s, NT=7s) (z = 3.25,p= 0.00056) in AR setup}
    \label{fig:ar_eye_contact_comparison}
\end{figure}
\begin{figure} [h]
    \centering
    \includegraphics[width=8cm,height=5cm]{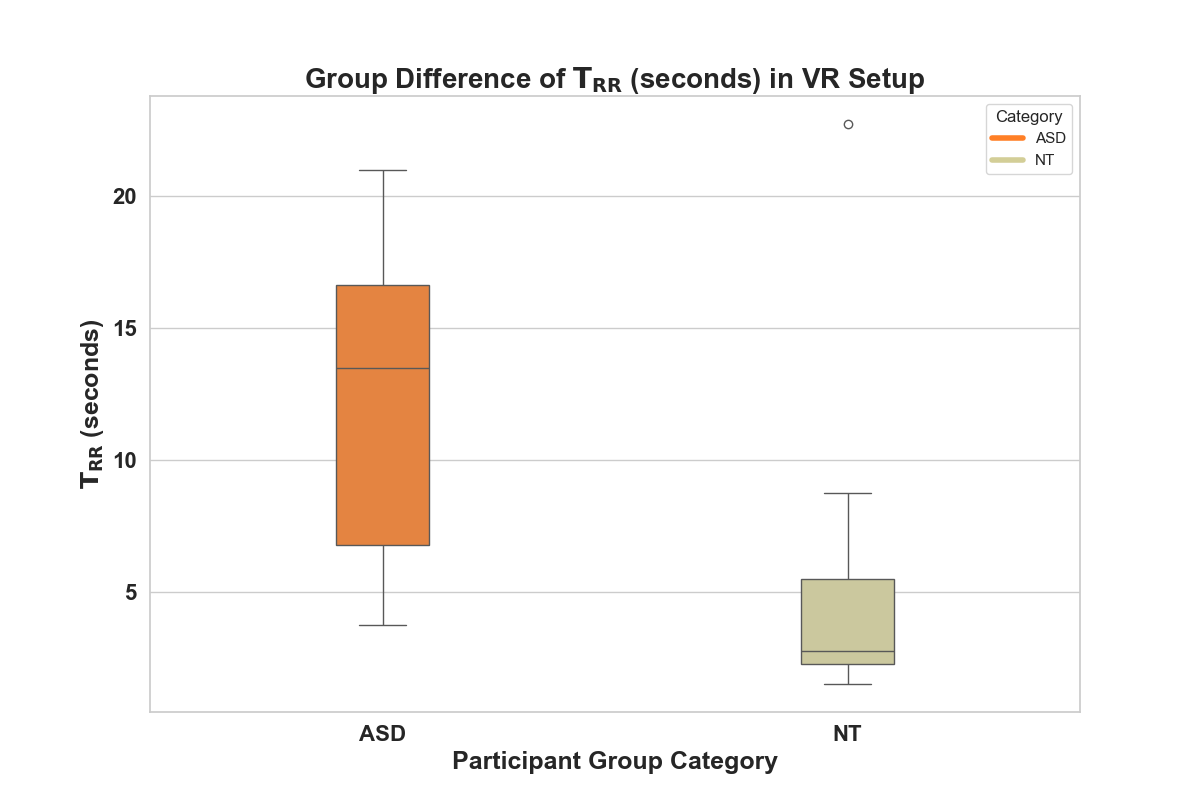}
    \caption{Duration to Register a response (T\textsubscript{RR}) comparison between ASD and NT participants (ASD=13.5s, NT=2.75s) (z=4.10,p=0.00074) in VR setup}
    \label{fig:vr_obj_registration_comparison}
\end{figure}
\begin{figure} [h]
    \centering
    \includegraphics[width=8cm,height=5cm]{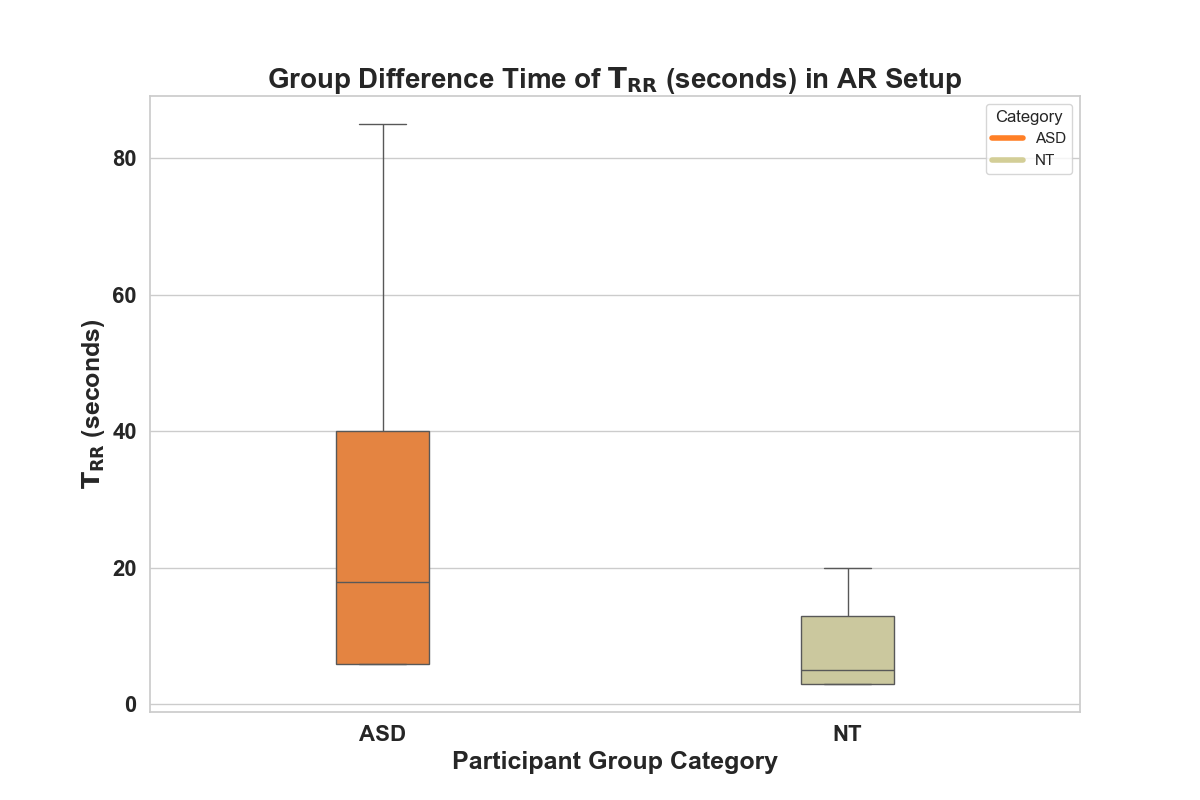}
    \caption{Duration to Register a response (T\textsubscript{RR}) comparison between ASD and NT participants (ASD=18s, NT=5s) (z=2.69,p=0.00357) in AR setup}
    \label{fig:ar_obj_registration_comparison}
\end{figure}

We used specific task performance metrics to compare and measure participant’s interaction with our JA training platforms on both the VR and AR setups. These metrics are (a) the correctness of the participant’s response (C\textsubscript{PR}), (b) the time taken by the participant to establish eye contact with the avatar (T\textsubscript{EC}), and (c) the time taken by the participant to register a response after the delivery of the cue (T\textsubscript{RR}). First, we analyze the group differences in the task performance of the participants. We utilised the non-parametric Mann-Whitney U test \cite{mann1947test} to analyze the relationship of the task performance metrics among the participant groups. This statistical one-tailed test was conducted at a significance level of 0.05.

\subsection{Statistical Group Differences of Task Performance Metrics}

We can observe from Table \ref{tab:groupDiffOfTaskPerformMetric} and Figures \ref{fig:vr_eye_contact_comparison} and \ref{fig:ar_eye_contact_comparison}, that the T\textsubscript{EC} of the NT participants (VR: median=3.75s; AR: median=7s) is significantly less (VR: p=0.00007; AR: p=0.00056) than their ASD counterparts (VR: median=11.5s; AR: median=35s). Similarly, from the Table \ref{tab:groupDiffOfTaskPerformMetric} and Fig \ref{fig:vr_obj_registration_comparison} and \ref{fig:ar_obj_registration_comparison}, we observe the T\textsubscript{RR} of the NT participants (VR: median=2.75s; AR: median=5s) is significantly less (VR: p=0.00074; AR: p=0.00357) compared to the ASD (VR: median=13.5s; AR: median=18s). 

It is possible that the higher T\textsubscript{RR} observed among ASD individuals during JA training might be due to their tendency to avoid eye contact, thereby leading to a prolonged response registration duration. 
The delayed response to the visual cue might result from difficulty in understanding the cue’s context and relevance rather than the difficulty in the cue processing.

\begin{figure} []
    \centering
    \includegraphics[width=8cm,height=5cm]{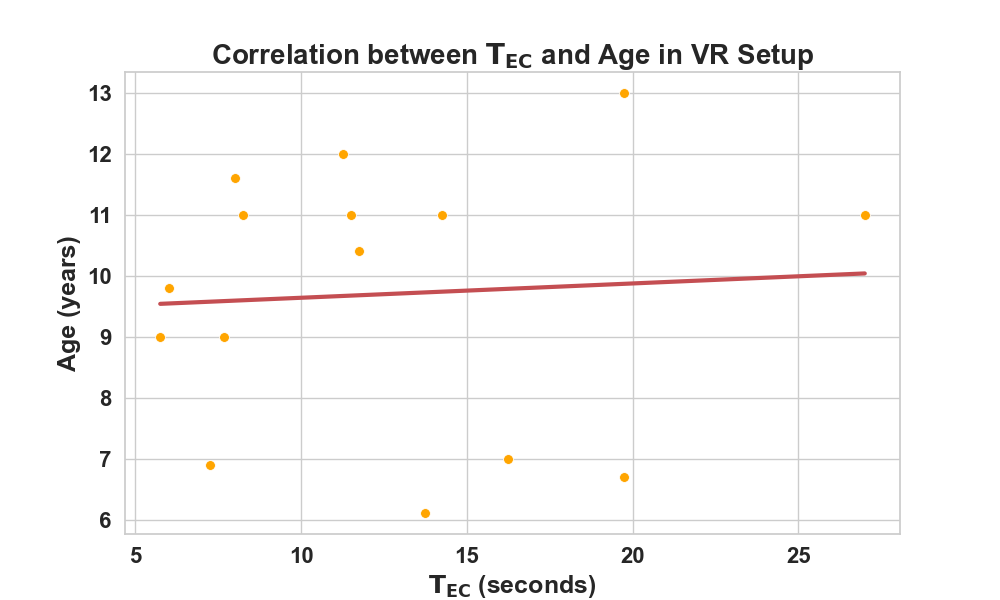}
    \caption{Correlation between Duration to Establish Eye Contact (T\textsubscript{EC}) and Age in ASD participants (rs=0.11,p=0.67) in VR setup}
    \label{fig:vr_asd_eye_contact_vs_age}
\end{figure}
\begin{figure} [h]
    \centering
    \includegraphics[width=8cm,height=5cm]{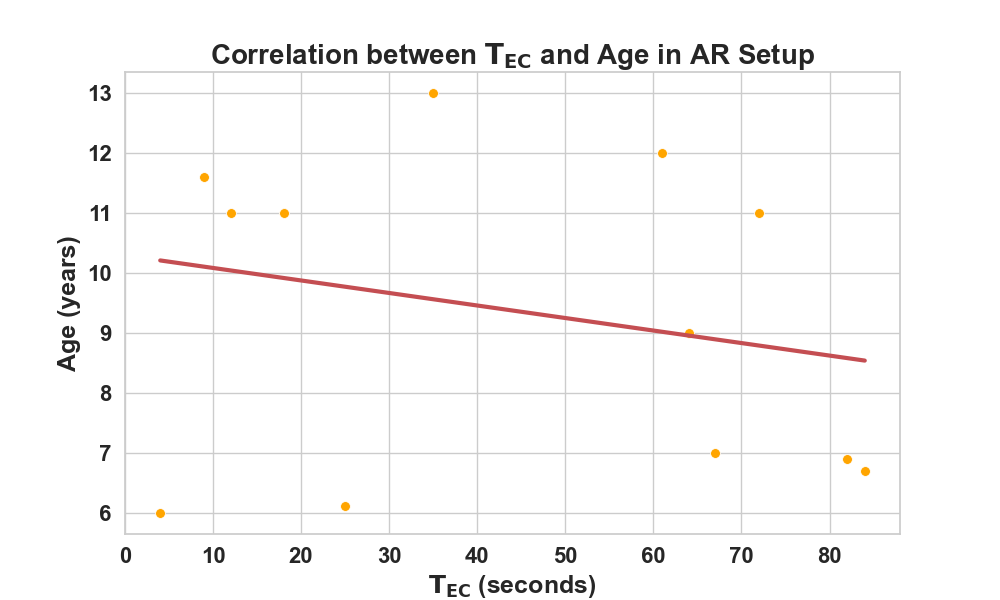}
    \caption{Correlation between Duration to Establish Eye Contact (T\textsubscript{EC}) and Age in ASD participants (rs=-0.15,p=0.60) in AR setup}
    \label{fig:ar_asd_eye_contact_vs_age}
\end{figure}
\begin{figure} [h]
    \centering
    \includegraphics[width=8cm,height=5cm]{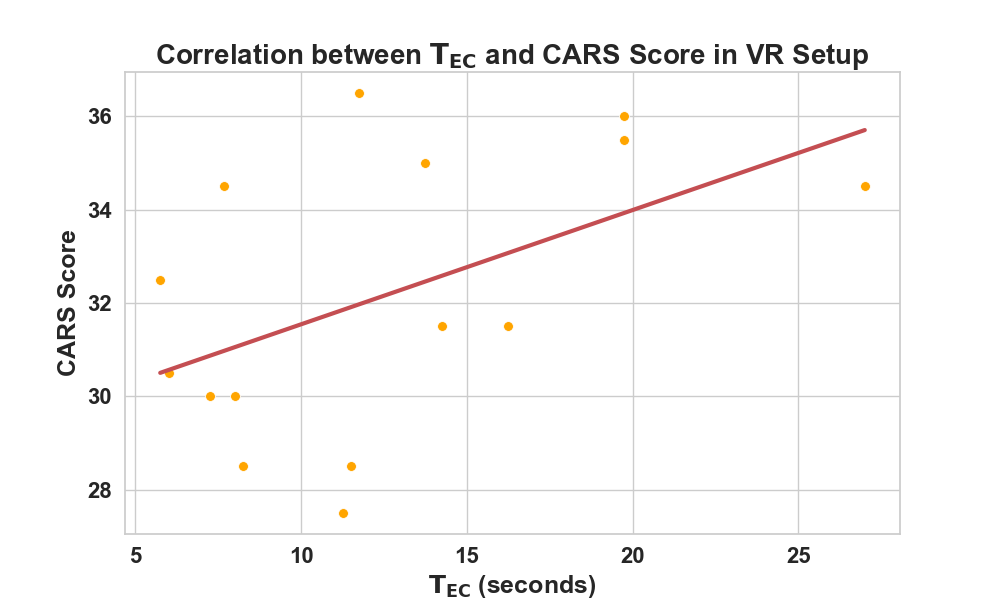}
    \caption{Correlation between Duration to Establish Eye Contact (T\textsubscript{EC}) and CARS score in ASD participants (rs=0.46,p=0.07) in VR setup}
    \label{fig:vr_asd_eye_contact_vs_cars}
\end{figure}
\begin{figure} [h]
    \centering
    \includegraphics[width=8cm,height=5cm]{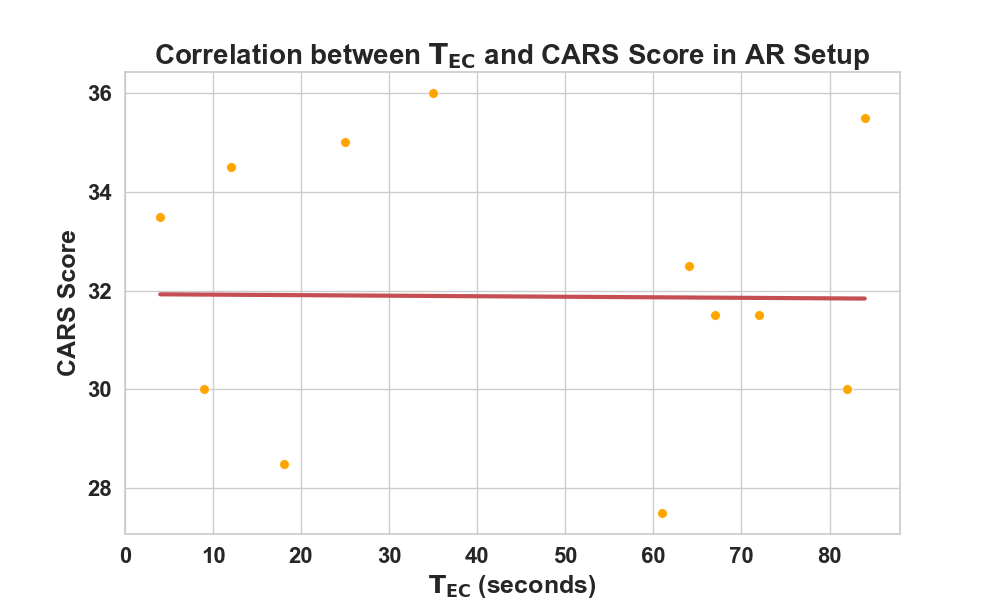}
    \caption{Correlation between Duration to Establish Eye Contact (T\textsubscript{EC}) and CARS score in ASD participants (rs=0.09,p=0.76) in AR setup}
    \label{fig:ar_asd_eye_contact_vs_cars}
\end{figure}
\begin{figure} [h]
    \centering
    \includegraphics[width=8cm,height=5cm]{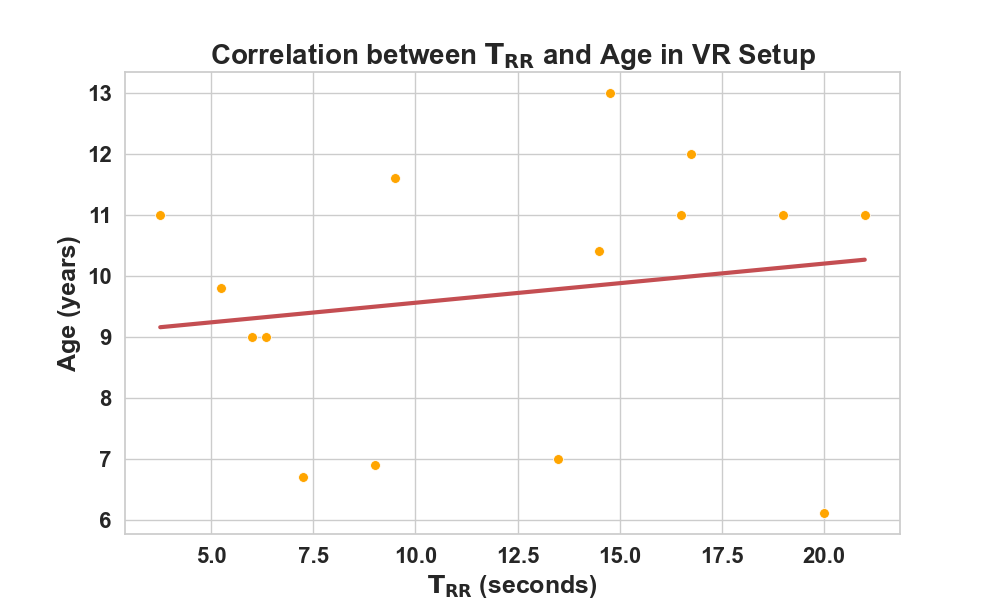}
    \caption{Correlation between Duration to Register a Response (T\textsubscript{RR}) and Age in ASD participants (rs=0.24,p=0.38) in VR setup}
    \label{fig:vr_asd_obj_registration_vs_age}
\end{figure} 
\begin{figure} [h]
    \centering
    \includegraphics[width=8cm,height=5cm]{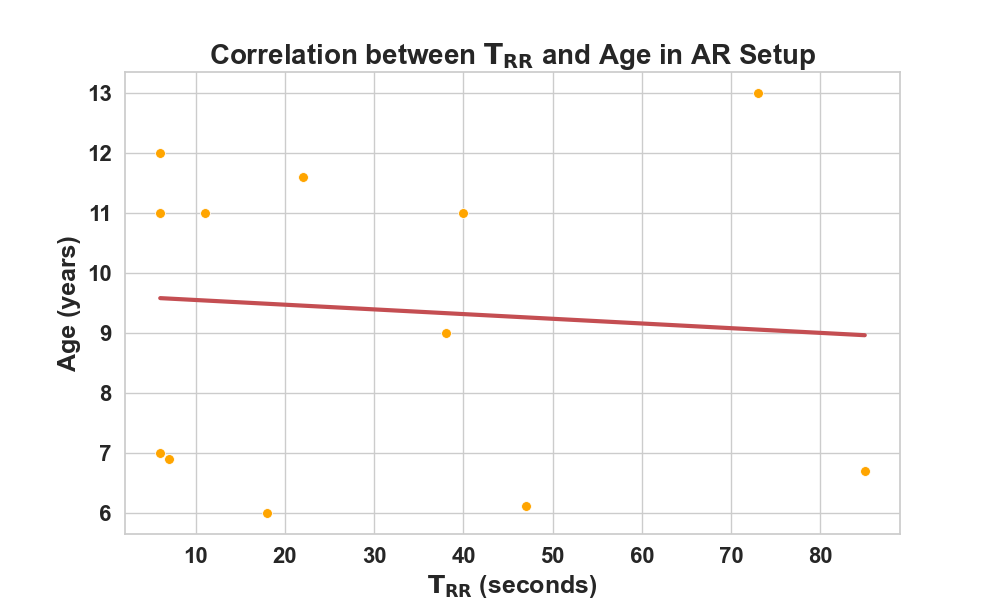}
    \caption{Correlation between Duration to Register a Response (T\textsubscript{RR}) and Age in ASD participants  (rs=-0.16,p=0.57) in AR setup}
    \label{fig:ar_asd_obj_registration_vs_age}
\end{figure} 
\begin{figure} [h]
    \centering
    \includegraphics[width=8cm,height=5cm]{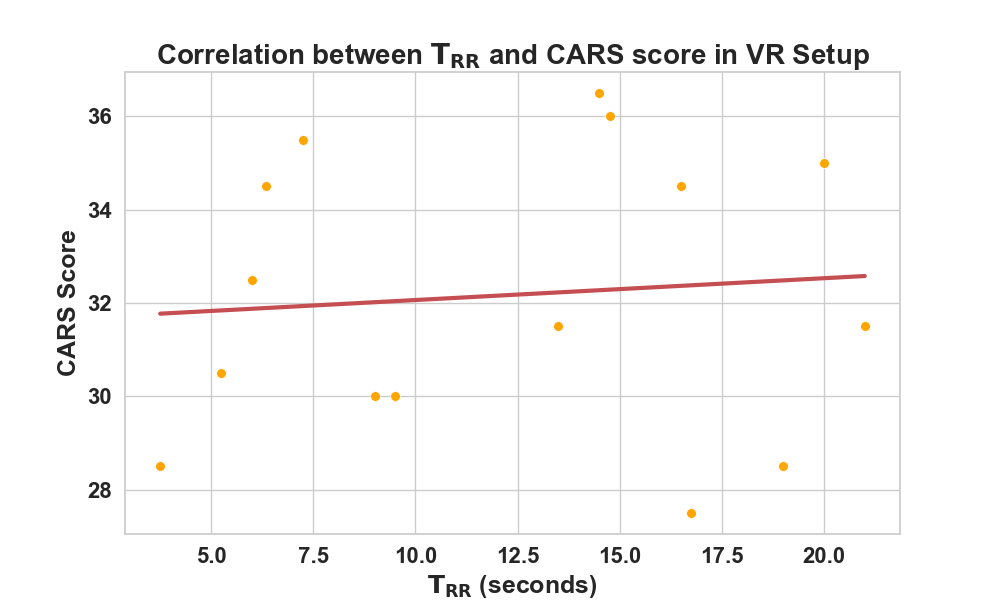}
    \caption{Correlation between Duration to Register a Response (T\textsubscript{RR}) and CARS score in ASD participants (rs=0.08,p=0.77)  in VR setup}
    \label{fig:vr_asd_obj_registration_vs_cars}
\end{figure}
\begin{figure} [h]
\vspace{-3mm} %
    \centering
    \includegraphics[width=8cm,height=5cm]{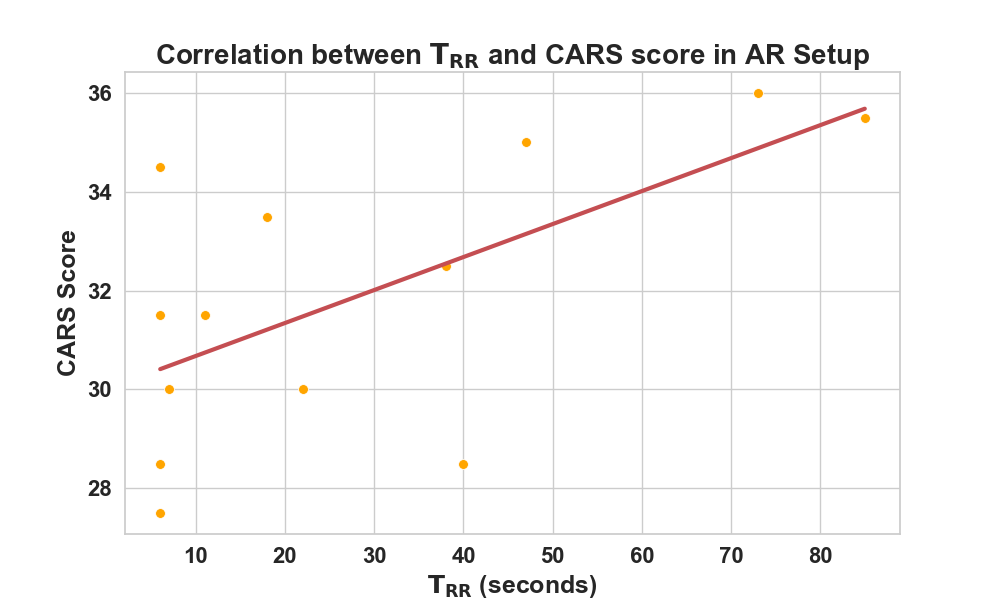}
    \caption{Correlation between Duration to Register a Response (T\textsubscript{RR}) and CARS score in ASD participants (rs=0.57,p=0.03) in AR setup}
    \label{fig:ar_asd_obj_registration_vs_cars}
\end{figure}

\begin{table}[h]
  \centering
  \caption{Comparison of Task Performance Metrics among ASD participants in VR and AR-based JA setups}
  \begin{center}
  \begin{tabularx}{8.8cm}{XXX}
    \toprule
     &   \textbf{\hspace{0.5cm} VR Setup} &   \textbf{\hspace{0.5cm} AR Setup} \\
         \cmidrule(lr){2-2} \cmidrule(lr){3-3} 
    Participants & \hspace{0.75cm} Mean & \hspace{0.75cm} Mean\\
    \midrule
    \hspace{0.25cm} T\textsubscript{EC} (s) & \hspace{0.75cm} 12.54 & \hspace{0.85cm} 42 \\
    \midrule
    \hspace{0.25cm} T\textsubscript{RR} (s) & \hspace{0.75cm} 12.20 & \hspace{0.85cm} 28 \\
    \midrule
    \hspace{0.25cm} C\textsubscript{PR} (\%) & \hspace{0.75cm} 69.5 & \hspace{0.85cm} 92.3 \\
    \bottomrule
  \end{tabularx}%
  \end{center}
  \label{tab:comparisonOfTaskPerformanceMetrics}%
  \footnotesize
\end{table}%
\begin{table}[h!]
    \centering
    \caption{Study of Statistical Group Differences of Task Performance Metrics among
ASD participants in VR and AR-based JA setups}
    \label{tab:grp_diff_analysis}
    \scriptsize
    \resizebox{\linewidth}{!}{
    \begin{tabular}{lcccccc}
        \toprule
        & \multicolumn{3}{c}{\textbf{VR Setup}} & \multicolumn{3}{@{\hskip -0.25 cm}c}{\textbf{AR Setup}} \\
        \cmidrule(lr){2-4} \cmidrule(lr){5-7}
        & \textit{p} & \textit{u} & \textit{z} & \textit{p} & \textit{u} & \textit{z} \\
        \midrule
        T\textsubscript{EC} & 0.00007 & 14.5 & 3.80039 & 0.00056 & 20.5 & 3.25641 \\
        \midrule
        T\textsubscript{RR} & 0.00074 & 28 & 3.17851 & 0.00357 & 31.5 & 2.69231 \\
        \bottomrule
    \end{tabular}
    }
        \caption*{\footnotesize {} Significance Level = 0.01, Hypothesis = 1 Tailed}
    \label{tab:groupDiffOfTaskPerformMetric} 
\end{table}
\begin{table}[h!]
    \centering
    \caption{Study of Correlation Analysis between Task Performance Metrics
and Participant Characteristics among ASD participants in VR and AR-based JA setups}
    \label{tab:corr_analysis}
    \begin{adjustbox}{width=0.5\textwidth}
    \footnotesize
    \begin{tabular}{lcccc}
        \toprule
        & \multicolumn{2}{c}{\textbf{VR Setup}} & \multicolumn{2}{c}{\textbf{AR Setup}} \\
        \cmidrule(lr){2-3} \cmidrule(lr){4-5}
        & \textit{p} & \textit{r} & \textit{p} & \textit{r} \\
        \midrule
        T\textsubscript{EC} vs CARS & 0.07840854 & 0.46816313 & 0.760216041 & 0.09392301 \\
        \midrule
        T\textsubscript{EC} vs Age & 0.67236623 & 0.119138426 & 0.603853662 & -0.159008343 \\
        \midrule
        T\textsubscript{RR} vs CARS & 0.775102028 & 0.080645679 & 0.038110133 & 0.57904873 \\
        \midrule
        T\textsubscript{RR} vs Age & 0.381882228 & 0.243474556 & 0.57987584 & -0.169491525 \\
        \bottomrule
    \end{tabular}
    
    \end{adjustbox}
    \caption*{\footnotesize {} r = Spearman's Rank Correlation Coefficient}
\end{table}

\subsection{Correlation Analysis between Task Performance Metrics and Participant Characteristics}

Further, we conducted a correlation analysis to examine the relationship between the task performance metrics (T\textsubscript{RR} and T\textsubscript{EC}) and the characteristics of the participants (age and CARS score). The CARS score acts as an indicator of the severity of autism, as it reflects the decline in an individual's social communication skills. The non-parametric Spearman's Rank Correlation Coefficient \cite{spearman1961proof} was utilized to determine the correlation between these parameters.
Our analysis shows that there was no significant correlation between the T\textsubscript{EC} and  T\textsubscript{RR} of the ASD participants with their CARS scores. 

Statistically, as observed from Fig. \ref{fig:vr_asd_eye_contact_vs_cars} and \ref{fig:ar_asd_eye_contact_vs_cars}, the p-value for T\textsubscript{EC} vs CARS for VR and AR is 0.07840 and 0.76021, respectively. Similarly, as shown in Fig. \ref{fig:vr_asd_obj_registration_vs_cars}, for T\textsubscript{RR} vs CARS, we obtained a p-value of 0.77510 for the  VR setup. For AR setup, as observed in Fig. \ref{fig:ar_asd_obj_registration_vs_cars}, we got a p-value of 0.03811. In contrast, in the above statistical comparisons, we observed a lack of correlation in most of the cases, which may be due to the limited range of CARS scores of the participants involved in the study. In just one instance we found a positive correlation, probably due to an increase in the response time of the participant with the CARS score.

Furthermore, from Fig. \ref{fig:vr_asd_eye_contact_vs_age} and  \ref{fig:ar_asd_eye_contact_vs_age}, we have also observed that there is no statistically significant correlation between the T\textsubscript{EC} (VR: p=0.67236; AR: p=0.60385). Similarly, from Fig. \ref{fig:vr_asd_obj_registration_vs_age} and \ref{fig:ar_asd_obj_registration_vs_age}, we observe the T\textsubscript{RR} (VR: p=0.38188; AR: p=0.57987) of the ASD participants and their age values. We believe this may be due to the limited age range of the participants involved in the study.

\subsection{Comparison of Correctness of Response (C\textsubscript{PR}) across the participant groups}
\begin{figure}[h!]
    \centering
    \includegraphics[width=8cm,height=6cm]{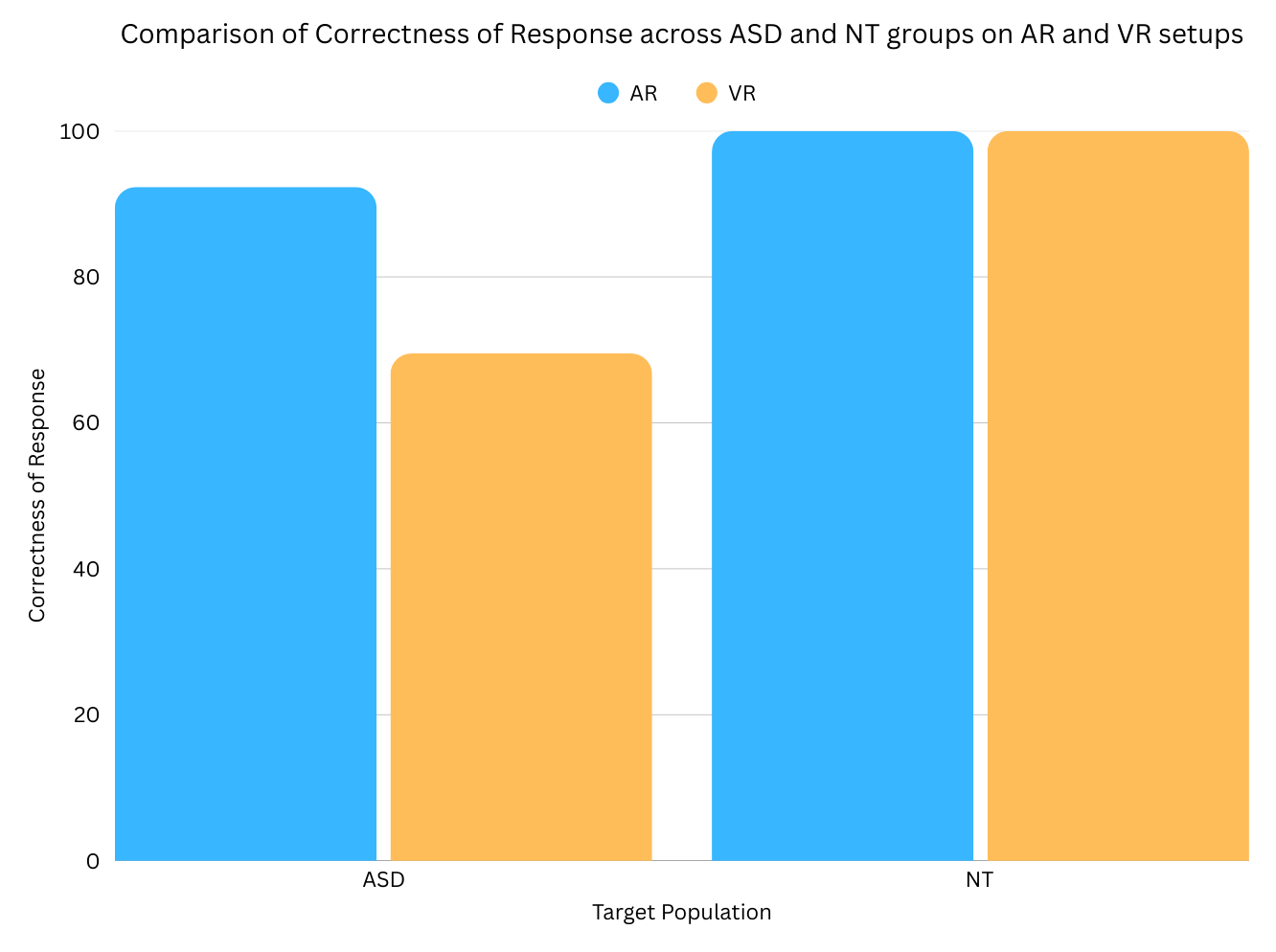}
    \caption{Comparison of Correctness of Response (C\textsubscript{PR}) for ASD and NT Participants across AR (ASD: 92.3\%, NT: 100\%) and VR (ASD: 69.5\%, NT: 100\%) setups}
    \label{fig:comparisonCorrectnessASDARVR}
\end{figure}

Finally, we performed a comparison of the C\textsubscript{PR}  among the ASD participants across both the VR and the AR setups. 
Upon analyzing the C\textsubscript{PR} of the ASD and NT groups on both the AR and VR setups, as shown in Fig. \ref{fig:comparisonCorrectnessASDARVR}, it was observed that 92.3\% of the participants in ASD correctly identified the target objects on the AR platform, while only 69.49\% performed the task correctly in the VR setup. This high accuracy can be attributed to the AR platform being more comfortable for ASD participants as they can still view the real-world environment, which helps them in coping with the social environment comfortably \cite{tarantino2023evaluation}. 

Nonetheless, as observed in Table \ref{tab:comparisonOfTaskPerformanceMetrics} the average T\textsubscript{EC} (AR:42s; VR:12.54s) and the T\textsubscript{RR} (AR:28s; VR:12.20s) values of ASD participants on the AR platform exceeded those on the VR platform. 
This behavior may be due to the fact that learning to interact with AR devices takes longer than with VR devices \cite{kyaw2023comparing}. This leaves us with an opportunity to improve the training of participants on the AR-based platform to obtain more optimal results in the future.
The study revealed that the participants with ASD exhibited a relatively high level of C\textsubscript{PR} for both the VR and AR platforms. This indicates that they can effectively comprehend the cues conveyed by the avatar and are capable of interacting with our JA training application. Our results corroborate with previous research studies that have utilized virtual avatars for JA training \cite{amat2021design,ravindran2019virtual,jyoti2019virtual,jyoti2020design,amaral2017novel}.
Overall, our solution can potentially serve as a reliable JA training platform.

\section{Discussion}

Early intervention to enhance JA abilities in children with ASD is widely recognized as critical, given its profound impact on long-term social communication and developmental outcomes. In this study, we present the development of immersive VR- and AR-based platforms designed to support JA skill training. Both platforms incorporate a digital avatar that functions as a task mediator, delivering visual cues to direct participants’ attention toward target objects in controlled environments.

We conducted a comparative feasibility study with 29 participants, employing a VR-based setup (FOVE 0 headset) and an AR-based setup (Microsoft HoloLens 2). The primary objective was to evaluate the feasibility, acceptability, and potential of these immersive systems for JA training in individuals with ASD. Both setups integrated eye-tracking modules, enabling precise, real-time measurement of participant gaze behaviors. A unified JA training platform was developed and subsequently adapted for deployment in both VR and AR devices. Task performance was quantified using three key metrics: time to establish eye contact (T\textsubscript{EC}), time to register a response following cue delivery (T\textsubscript{RR}), and correctness of response (C\textsubscript{PR}).

Our results demonstrate that neurotypical participants exhibit significantly shorter T\textsubscript{EC} and T\textsubscript{RR} compared to their ASD counterparts across both VR and AR setups. Furthermore, correlation analysis reveals a significant positive association between T\textsubscript{RR} and CARS scores among ASD participants, suggesting that response latency increases with autism severity. Interestingly, this trend is more evident in the AR setup than in the VR platform. One possible explanation is that AR offers a more relatable and engaging environment for individuals with ASD, whereas fully immersive VR may be experienced as overwhelming or less ecologically familiar. No significant correlations are observed between task performance and participant age across either setup, which may be attributable to the limited sample size. 

Nevertheless, both VR and AR systems demonstrate consistently high C\textsubscript{PR} values, indicating that participants accurately interpret the avatar’s visual cues and effectively employ gaze-based interactions to identify target objects. These findings suggest that the platforms successfully capture and retain participant attention, a critical factor for training efficacy.

Overall, this study highlights the feasibility and potential of immersive VR and AR-based systems as innovative tools for JA intervention. The high accuracy rates observed across both platforms indicate that children with ASD are able to engage meaningfully with avatar-mediated cues, while differential trends in response latency underscore the importance of tailoring immersive environments to individual needs.

\section{Conclusion}
This study presents the design, development, and evaluation of an immersive JA training platform implemented on both VR and AR devices, incorporating avatar-mediated visual cues and integrated eye tracking technology. Our comparative feasibility study engaged children diagnosed with ASD alongside age-matched NT participants. The results indicate that the platform effectively captures gaze-based interactions, delivers consistent training stimuli, and maintains participant engagement. The findings reveal that NT participants demonstrate significantly faster gaze responses compared to their ASD counterparts. However, both groups achieved high correctness scores, reflecting an accurate interpretation of avatar cues. Additionally, there is a significant association between autism severity and response latency, which emphasizes the platform’s potential for capturing clinically meaningful behavioral markers. Importantly, this study suggests that AR-based interventions may provide a more accessible and less overwhelming experience for individuals with ASD when compared to fully immersive VR environments. 

Collectively, these results offer preliminary evidence supporting the feasibility, acceptance, and promise of immersive JA training tools as innovative strategies for early intervention in ASD. While further validation with larger and more diverse samples is essential, as well as longitudinal investigations and the integration of multimodal cues, this work establishes a vital foundation for the future development of personalized, technology-driven interventions aimed at enhancing core social-communicative skills in children with ASD.

\section{Limitations and Future Works}

To the best of our knowledge, this study represents the first attempt to develop an avatar-mediated JA training platform optimized for AR devices. The positive acceptance demonstrated by participants provides preliminary evidence of the feasibility and potential of this approach.

Nevertheless, the study has several limitations. The relatively small sample size restricts the generalizability of our findings and limits the strength of conclusions regarding the comparative effectiveness of VR and AR setups. Future studies with larger and more diverse cohorts are required to validate these results and to enable more detailed analyses across participant subgroups. We also plan to enhance the training paradigm by incorporating verbal cues alongside visual cues, thereby aligning the platform more closely with real-world communicative contexts. Furthermore, instead of restricting analysis to predefined regions of interest (ROIs), we intend to track participants’ full fixation patterns throughout the task. Such data will yield deeper insights into attentional strategies and facilitate the design of more refined and adaptive JA training exercises.
Finally, we recognize the need for longitudinal studies to assess the sustained impact of the platform on JA skill development in children with ASD. It will be equally important to evaluate the ecological validity of these interventions by examining how the skills acquired in immersive environments translate into real-world social interactions, which are inherently more complex and nuanced than simulated scenarios.

\IEEEpeerreviewmaketitle


\ifCLASSOPTIONcaptionsoff
  \newpage
\fi

\bibliographystyle{IEEEtran}
\bibliography{bib}

\begin{IEEEbiography}[{\includegraphics[width=1in,height=1.25in,clip,keepaspectratio]{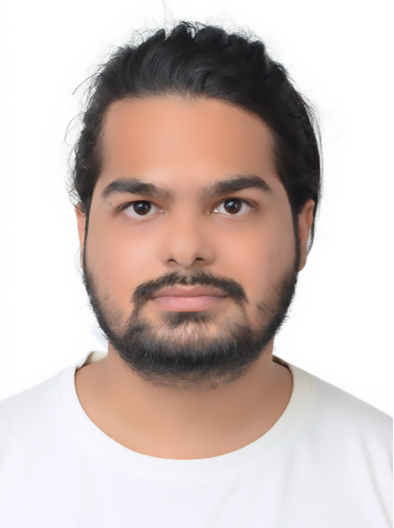}}]{Ashirbad Samantaray}
received the MS Research degree in Electrical Engineering (Computer Technology group) from Indian Institute of Technology Delhi in 2024. He is currently a Research Engineer at iHub-Drishti Foundation, IIT Jodhpur. His research interests include Virtual Reality, Augmented Reality, Autism and Human Computer Interaction.
\end{IEEEbiography}
\begin{IEEEbiography}[{\includegraphics[width=1in,height=1.25in,clip,keepaspectratio]{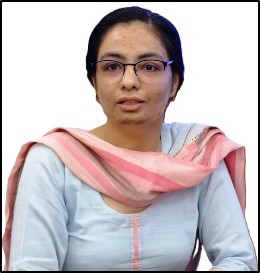}}]{Taranjit Kaur} received the Ph.D. in Electronics and Communication Engineering from Dr. B. R. Ambedkar National Institute of Technology, Jalandhar, Punjab, in 2017.  

Currently, she is a Scientific Officer at the School of AI and Data Sciences at the Indian Institute of Technology Jodhpur. She specializes in the areas of computational neuroscience, Artificial Intelligence, and Assistive technology.
\end{IEEEbiography}
\begin{IEEEbiography}[{\includegraphics[width=1in,height=1.25in,clip,keepaspectratio]{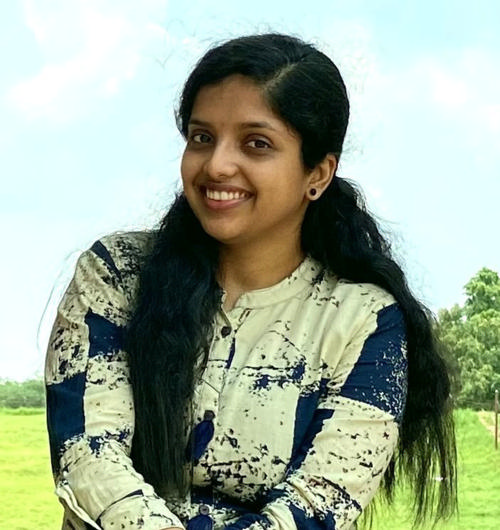}}]{Sapna S Mishra}
received the PhD degree in Electrical Engineering from Indian Institute of Technology Delhi in 2025. She is currently working as a Postdoctoral researcher in New Jersey Institute of Technology, New Jersey. Her research interests include Brain Imaging, Autism and Human Computer Interaction.
\end{IEEEbiography}
\begin{IEEEbiography}[{\includegraphics[width=1in,height=1.25in,clip,keepaspectratio]{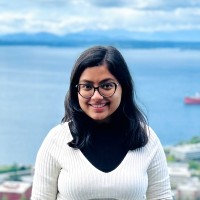}}]{Kritika Lohia}
received the PhD degree in Electrical Engineering from Indian Institute of Technology Delhi in 2025. She is currently working as a Postdoctoral researcher in University of St Andrews, Scotland. Her research interests include Human Visual systems, Autism and Human Computer Interaction.
\end{IEEEbiography}
\begin{IEEEbiographynophoto}{Chayan Majumder}
Chayan Majumder received the Master of Technology degree from BITS Pilani, India and Bachelor of Technology degree from the University of Delhi, India in Electronic and Communication Engineering.
He is currently a research scholar pursuing his  at the Indian institute of technology Delhi. 
Area of research is computational neuroscience.
\end{IEEEbiographynophoto}
\begin{IEEEbiography}[{\includegraphics[width=1in,height=1.25in,clip,keepaspectratio]{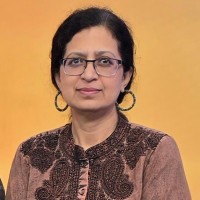}}]{Sheffali Gulati}
received the MD degree from All India Institute of Medical Sciences Delhi (AIIMS Delhi) in 1998. She is currently a Professor at the Child Neurology Division at AIIMS, New Delhi. Her research interests include Pediatric Neurology, Autism and Epilepsy. She is a member of International Society for Autism Research and currently serves as the president of the Association of Child Neurology India.
\end{IEEEbiography}
\begin{IEEEbiography}[{\includegraphics[width=1in,height=1.25in,clip,keepaspectratio]{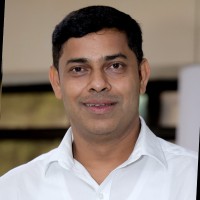}}]{Tapan Kumar Gandhi}
received the PhD degree from Indian Institute of Technology Delhi (IITD) in 2012. He is currently working as a Professor at the Department of Electrical Engineering IITD. His research interests include Healthcare Technology, Computational Neuroscience, Autism and Human Computer Interaction. He is elected Fellow of both National Academy of Engineering (FNAE) and National Academy of Sciences (FNASc.).
\end{IEEEbiography}

\end{document}